\documentclass[showpacs,aps,prd,preprint,nofootinbib,showkeys,unsortedaddress,raggedbottom]{revtex4-1}
\pdfoutput=1

\usepackage{bm}
\usepackage{amsmath}
\usepackage{graphicx}
\usepackage{subfigure}
\usepackage{float}
\usepackage[usenames,dvipsnames]{color}
\definecolor{darkblue}{RGB}{0,0,196}
\usepackage[colorlinks=true,linkcolor=darkblue,citecolor=darkblue,urlcolor=darkblue]{hyperref}

\usepackage{setspace}
\usepackage{footmisc}
\usepackage[makeroom]{cancel}

\def\be{\begin{equation}}
\def\ee{\end{equation}}
\def\ba{\begin{eqnarray}}
\def\ea{\end{eqnarray}}

\begin{document}

\title{Parton self-energies for general momentum-space anisotropy}

\author{Babak S. Kasmaei and Michael Strickland}
\affiliation{Department of Physics, Kent State University, Kent, OH 44242 United States}

\begin{abstract}
We introduce an efficient general method for calculating the self-energies, collective modes, and dispersion relations of quarks and gluons in a momentum-anisotropic high-temperature quark-gluon plasma. The method introduced is applicable to the most general classes of deformed anisotropic momentum distributions and the resulting self-energies are expressed in terms of a series of hypergeometric basis functions which are valid in the entire complex phase-velocity plane.  Comparing to direct numerical integration of the self-energies, the proposed method is orders of magnitude faster and provides results with similar or better accuracy.  To extend previous studies and demonstrate the application of the proposed method, we present numerical results for the parton self-energies and dispersion relations of partonic collective excitations for the case of an ellipsoidal momentum-space anisotropy. Finally, we also present, for the first time, the gluon unstable mode growth rate for the case of an ellipsoidal momentum-space anisotropy.  
\end{abstract}

\maketitle

\section{Introduction}

The evidence for the generation and evolution of a new state of matter called the quark-gluon plasma (QGP) has been provided by heavy-ion collision experiments performed at RHIC and LHC.  There are many interesting questions about the thermodynamics, transport properties, structure of the phase diagram, and the characteristics of the phase transitions at the various evolution stages of such an extreme but short-lived system~\cite{Muller:2012zq,Jacak-Muller:2012, Schukraft:2017nbn}.  The interactions of quark and gluon degrees of freedom play the main role in the realization of the properties of the QGP and its evolution. Non-perturbative aspects of the strong and intermediate coupling regimes of quantum chromodynamics (QCD) make it difficult to be used directly for the phenomenological studies of heavy-ion collisions and QGP. As a result, the current theoretical understanding of QGP is mostly based on lattice QCD \cite{Ding:2015ona}, hard-(thermal)-loop perturbation theory \cite{Braaten:1989mz,Andersen:2004fp,Haque:2014rua}, effective hydrodynamic models \cite{Gale:2013da,Jeon:2016uym,Alqahtani:2017mhy}, AdS/CFT inspired dual descriptions of the QGP~\cite{Son:2007vk,casalderrey2014gauge}, etc.

In the course of the last decades, researchers have come to learn that the QGP, as created in the heavy-ion collisions experiments, is most likely neither in perfect kinetic nor chemical equilibrium. In particular, one finds that the deviations from momentum-space isotropy are quite large, especially at early times after the nuclear impact and/or in the dilute regions of the QGP \cite{Strickland:2013uga,Alqahtani:2017mhy}.  Hard-loop perturbation theory can be generalized to incorporate such non-equilibrium effects, where an effective phase-space description of a semi-classical partonic medium can be provided by considering certain classes of non-equilibrium momentum distribution functions \cite{Mrowczynski:2004kv}. In this framework, the non-equilibrium dynamics of a non-equilibrium plasma can be described in terms of collective excitations of the gluon and quark degrees of freedom, namely color plasmon and plasmino quasiparticles. 

The dispersion relations for the non-equilibrium collective modes of QGP can be deduced from the singularities of the dressed propagators of the partonic quasiparticles. Their self-energies modify their bare masses due to medium effects within the plasma and, in a momentum-space anisotropic QGP, depend explicitly on their direction of propagation. The self-energies and dispersion relations for QGP partons possessing a generic anisotropy of momentum distributions has been calculated by Mr\'owczy\'{n}ski and Thoma, where they showed the equivalence of the results obtained from kinetic transport and hard loop effective theories \cite{Mrowczynski:2000ed}.\footnote{This was done at leading order in the strong coupling constant with the only requirement being that the distribution function be reflection-symmetric in momentum-space, $f({\bf p}) = f(-{\bf p})$.} Subsequently, Romatschke and Strickland introduced a convenient parameterization of the momentum anisotropy of the QGP using an anisotropic deformation of the argument of the isotropic (equilibrium) distribution functions \cite{Romatschke:2003ms,Romatschke:2004jh}. Using this parameterization, the collective modes of a QGP with a spheroidal momentum anisotropy with one anisotropy parameter were calculated for both gluons \cite{Romatschke:2003ms, Romatschke:2004jh, Carrington:2014} and quarks \cite{Schenke:2006}.  It was shown in these prior works that, in a momentum-space anisotropic QGP, the gluon excitations possess unstable modes related to the so-called chromo-Weibel instability \cite{Mrowczynski:2016etf}. This description has been also used to study various aspects of a momentum-space anisotropic QGP.  In particular, the effects of unstable modes on the thermalization and isotropization of the QGP \cite{Mrowczynski:2016etf}, photon and dilepton production from the QGP \cite{Schenke:2007, Bhattacharya:2016}, the QGP heavy quark potential \cite{Nopoush:2017zbu}, and bottomonia suppression \cite{Krouppa:2016jcl} have been previously studied, for example. In addition, the introduction of anisotropic momentum distribution functions to relativistic hydrodynamic models has provided a successful description of QGP evolution between the hydronization and hadronization stages \cite{Alqahtani:2017jwl,Alqahtani:2017mhy}. 

Previous studies of the collective phenomena in a momentum-space anisotropic QGP have mostly addressed the special case of spheroidal anisotropy specified by one direction and one anisotropy parameter. However, additional effects are expected to arise in more general cases and the introduction of additional anisotropy parameters is necessary to provide a better characterization of QGP properties, evolution, and observables. In particular, calculating the photon production rates of a QGP with ellipsoidal momentum anisotropy, where two independent anisotropy parameters are considered, might be useful in characterization of the elliptic flow of photons and dileptons. In a previous paper, the calculation of the quark self-energy for an ellipsoidally anisotropic QGP was addressed \cite{Kasmaei:2016apv}. However, most of the methods for those calculations have been case-specific, time-consuming, and even in some cases intractable. In this paper, we introduce an efficient and general method for the calculation of quark and gluon self-energies which is applicable to all relevant forms, strengths of momentum anisotropy, propagation directions, and complex phase velocities. 

This paper is organized as follows: The description of a momentum-anisotropic QGP, the quark and gluon hard-loop self-energies in such a system, and the parameterization of the momentum anisotropic distribution functions are addressed in Sec.~\ref{aqgp}, where a generalized approach to the anisotropy parameterization also extends the prior literature.  In Sec.~\ref{frame}, two choices of plasma and parton coordinate systems are introduced, where the simplicity of the parton frame for calculational purposes is demonstrated. In Sec.~\ref{methods} an efficient and general method for the calculation of the self-energy integrals is proposed which uses the information contained solely in the imaginary part of the self-energy in a finite real interval to extract both the real and imaginary parts in the entire complex plane as a weighted sum of hypergeometric functions. The piecewise version of this method is also addressed. The results for the quark and gluon self-energies and dispersion relations in an ellipsoidally anisotropic QGP, calculated using the proposed method, are presented in Sec.~\ref{results}. The dispersion curves of the real stable modes for quarks and gluons and the imaginary unstable modes of the gluons are plotted and compared to the spheroidally-anisotropic case. The concluding remarks and an outlook for future studies are given in Sec.~\ref{conclusion}. In App.~\ref{appndxa}, a method to obtain analytic results for the imaginary parts of the self-energy is presented and an example analytic result is given. Various properties of the basis functions used in the body of the text and their recursive relations etc. are presented in App.~\ref{appndxb}.

\section{Momentum anisotropic quark-gluon plasma}
\label{aqgp}

The propagator of a quark or gluon with four-momentum $K=(\omega, {\bf k})$ in a hot and anisotropic medium is modified by interactions with the massless partonic degrees of freedom with four-momentum $P=(p, {\bf p})$.  The hard-loop gluon  self-energy (polarization) tensor in a non-equilibrium QGP can be written as~\cite{Mrowczynski:2000ed}
\be
\Pi^{\mu\nu}(k)= g^2 \int \frac{d^3 \bf p}{(2 \pi)^3} \, \frac{p^\mu}{E_{\bf p}} \, \frac{\partial  f(\bf p)}{\partial p^\lambda}  \left[\eta^{\lambda \nu} - \frac{k^\lambda p^\nu}{p^\sigma k_\sigma + i 0^{+}} \right ], \label{gse0} 
\ee
where $g$ is the strong coupling constant, $\eta^{\lambda \nu} = {\rm diag}(1,-1,-1,-1)$ is the Minkowski space metric tensor, and $f({\bf p})$ is the effective one-particle distribution function which will be specified later.  The gluon self-energy tensor is transverse to the gluon momentum
\be
k_\mu \Pi^{\mu \nu}=0 \, , 
\ee
and, as a result, its space-like components can be used to construct the full four-tensor. Assuming massless partons  ($E_{\bf p}=|\bf p|$), the spatial part of this tensor can be written as
\be
\Pi^{ij}(k)= -g^2 \int \frac{d^3 \bf p}{(2 \pi)^3} v^i \ \frac{\partial  f(\bf p)}{\partial p_l} \left[\delta^{j l} + \frac{k^l v^j}{\omega -{\bf k . v}  +i 0^{+}} \right] . \label{gse1} 
\ee

Similarly, the hard-loop quark self-energy in a non-equilibrium QGP is~\cite{Mrowczynski:2000ed}
\ba 
 \Sigma(K) = \frac{g^2 C_F}{4}  \int  \frac{d^3 \bf p}{(2 \pi)^3} \frac{f ({\bf p})}{|{\bf p}|} \  \frac{\gamma_0 + {\bf v.\boldsymbol \gamma}}{\omega -{\bf k.v} + i0^{+}} \, , \label{qse0} 
\ea
where the $\gamma_{\mu}=(\gamma_0, \boldsymbol \gamma)$ are the Dirac matrices and the color factor $C_F = (N_c^2 -1)/(2N_c)$ is the quadratic Casimir in the fundamental representation.  We note that, in this context, it is useful to define the components of a quark self-energy four-vector $\Sigma^{\mu}=(\Sigma^0, \boldsymbol \Sigma)$ with $\displaystyle \Sigma = \gamma_{\mu}\Sigma^{\mu}$.

In the self-energy integrals \eqref{gse1} and \eqref{qse0}, the effective non-equilibrium parton momentum distribution functions are
\be
f({\bf p})=
\left\{
    \begin{array}{lr}
  2 N_c  n_g ({\bf p}) + N_f \big[ n_q({\bf p}) + n_{\bar{q}}({\bf p}) \big]  & \text{for the gluon self-energy integral} \, , \\
    2 n_g ({\bf p}) + N_f \big[ n_q({\bf p}) + n_{\bar{q}}({\bf p}) \big] & \text{for the quark self-energy integral} \, ,
     \end{array}
     \right.
\ee   
where $n_g ({\bf p}) , n_q({\bf p})$, and $n_{\bar{q}}({\bf p})$ are gluon, quark, and antiquark number densities, respectively, and $N_c$ and $N_f$ are the number of colors and quark flavors, respectively.

A general method for introducing anisotropic momentum distribution functions $f({\bf p})$ is to make a transformation of the argument of an isotropic momentum distribution $f_{\rm iso}(p=|{\bf p|)}$ to deform them into
\be
f_{\{ {\boldsymbol \alpha}\}}({\bf p})= f_{\rm iso}\!\left(\frac{|{\bf p}|}{\Lambda(\{ {\boldsymbol \alpha}\})} \, H\big({\bf \hat{v}}; \{{\boldsymbol \alpha}\} \big) \right) \, ,
\label{anisodistgen}
\ee
where $\Lambda$ is a temperature-like momentum scale, the anisotropic transformation function $H\big({\bf \hat{v}}; \{{\boldsymbol \alpha}\} \big)$ depends on the momentum unit vector ${\bf \hat{v}} $ in spherical coordinates confined to the surface of a unit-sphere $\Omega \equiv  (\theta , \phi)$ and we have introduced a set of parameters $\{{\boldsymbol \alpha}\}$ which specify the form and strength of the anisotropy.  In Eq.~\eqref{anisodistgen} we have indicated that $\Lambda$ can depend on the parameters $\{{\boldsymbol \alpha}\}$ in order to allow for generalized matching of the energy density, number density, etc.~if required \cite{Martinez:2010sc}.  When it is clear, we will omit the specification of the set $\{ {\boldsymbol \alpha}\}$.  The function $H({\bf \hat{v}})$ should return a real-valued positive number in order to guarantee that the argument of $f_{\rm iso}(x)$ is positive and that the distribution function is real valued.  For a direction ${\bf \hat{v}}$,  this transformation induces an expansion of the iso-surfaces of the isotropic distribution for that direction if $0<H({\bf \hat{v}}) <1$, or induces their contraction if  $H({\bf \hat{v}})>1$.

One widely used special case of the transformation above is that of a spheroidal momentum deformation introduced originally by Romatschke and Strickland \cite{Romatschke:2003ms}.  It is defined using a single anisotropy direction $\bf \hat{n}$ and parameter $\xi$ as
\be
H_s({\bf \hat{v}};\{\xi,{\bf \hat{n}} \}) = \sqrt{1+ \xi ({\bf \hat{n} \cdot \hat{v}})^2} \, . \label{spheroidal}
\ee
For a QGP with an ellipsoidal momentum anisotropy and two independent anisotropy directions $\bf \hat{n}_1$ and $\bf \hat{n}_2$, one can generalize \eqref{spheroidal} to
\be
H_e({\bf \hat{v}};\{\xi_1,\xi_2,{\bf \hat{n}_1},{\bf \hat{n}_2} \}) = \sqrt{1+ \xi_1 ({\bf \hat{n}_1\cdot\hat{v}})^2 + \xi_2 ({\bf \hat{n}_2\cdot\hat{v}})^2} \, .  \label{ellipsoidal}
\ee  
This can be further generalized to a quadratic form, also known as the generalized Romatschke-Strickland parameterization, where a tensor $\Xi^{\mu\nu}$ defines the anisotropic transformation~\cite{Martinez:2012tu,Nopoush:2014pfa}
\be
H_\Xi({\bf \hat{v}};\Xi^{\mu\nu})  = \sqrt{v^{\mu}\Xi_{\mu\nu}v^{\nu} } \qquad \text{with} \ \ v^{\mu}=(1,{\bf \hat{v}}) \, . \label{quadratic}
\ee

With a general anisotropy function $H(\theta, \phi) >0$, the quark self-energy \eqref{qse0} takes the form
\be
 \Sigma(K) = m_q^2 \  \int  \frac{d \Omega}{4\pi} \ \frac{1}{\left[H(\theta, \phi)\right]^2} \  \frac{\gamma_0 + {\bf \hat{v}\cdot\boldsymbol \gamma}}{\omega -{\bf \hat{k}\cdot\hat{v}} + i0^{+}} \, , \label{qse2}
\ee
with the quark effective mass defined by
\be
m_q^2= \frac{g^2 C_F}{8 \pi^2} \int_0^{\infty} dp \, p \, f_{\rm iso}(p/\Lambda) \, . \label{mq2}
\ee

For a quadratic momentum anisotropy defined by $H_\Xi(\theta,\phi)$, we can also rewrite the components of the gluon polarization tensor \eqref{gse1} as\,\footnote{We have yet to find an equivalent expression for a general anisotropic deformation.}
\be
\Pi^{ij}(K) =\ m_D^2 \int \frac{d \Omega}{4\pi} v^i  \ \frac{ W^l}{ \left[H_\Xi(\theta,\phi) \right]^4} \left[\delta^{j l} + \frac{k^l v^j}{\omega -{\bf \hat{k} \cdot \hat{v}}  +i 0^{+}} \right] , \label{gse2}
\ee
with the gluon Debye mass defined as
\be
m_D^2 = - \frac{g^2}{2 \pi^2} \int_0^{\infty} dp \, p^2 \, \frac{df_{\rm iso}(p/\Lambda)}{dp} \, , \label{md2}
\ee
and
\be
W^l = \left(  \Xi^{00}+  \Xi^{ll}+\frac{\Xi^{0l}+\Xi^{l0}}{2}v^l \right) v^l + \frac{\Xi^{0l}+\Xi^{l0}}{2}  + \sum_{r \neq l}\frac{\Xi^{rl} + \Xi^{lr}}{2} v^r \, .
\ee

\section{Plasma and parton coordinate systems}
\label{frame}

In order to calculate the integrals \eqref{qse2} and \eqref{gse2} for general $\bf \hat{v}$, one needs to specify an appropriate choice of coordinate system. One choice is to define the polar angle $\theta$ by $x=\cos \theta = {\bf \hat{k} \cdot \hat{v}}$ and the azimuthal angle $\phi$ in the plane defined by $\theta=\pi/2$. In this case, the integration variables used to calculate the self-energies of a parton are defined relative to the direction of the parton momentum. The benefit of this definition is that it becomes straightforward to see that the integrand has a single first-order pole at $x = z + i\epsilon$ where $z \equiv \omega/k$ is the phase space velocity (speed).  In practice, for an expanding QGP one may consider a time-varying description of the anisotropic momentum distributions defined for the whole or a local region of the plasma. Then, it is useful to rotate the calculated self-energies from parton-specific coordinates to the plasma coordinates. 

For the case of ellipsoidal anisotropy, if we consider ${\bf \hat{n}_1}'=(0,0,1)$, ${\bf \hat{n}_2}'=(1,0,0)$, and ${\bf \hat{k}}' = (\sin \theta_k \cos \phi_k, \ \sin \theta_k \sin \phi_k, \ \cos \theta_k)$ to define the anisotropy and parton momentum vectors in the plasma primed coordinates, then in the coordinates defined for that parton we have
\ba
&{\bf \hat{k}}&\ =(0,0,1) \, , \\
&{\bf \hat{n}_1}&\ = (0, \ -\sin \theta_k, \ \cos \theta_k) \, , \\
&{\bf \hat{n}_2}&\ = (\sin \phi_k, \ \cos \theta_k \cos \phi_k, \ \sin \theta_k \cos \phi_k) \, , \\
&{\bf \hat{v}}&\ = (\sin \theta \cos \phi, \ \sin \theta \sin \phi, \ \cos \theta) \, .
\ea

\section{Calculation method for anisotropic self-energies}
\label{methods}

As mentioned in the previous section, the pole of the integrand of the self-energies \eqref{gse1} and \eqref{qse0} has the simplest form in the coordinate system specified by the external parton, where the polar angles of the internal hard-loop partons are defined relative to the direction of external parton momentum $\bf k$. Since $H(x\equiv\cos \theta, \phi)$ has no zeros on the surface of the unit sphere defined by $(x,\phi)$, using the Sokhotski-Plemelj theorem \cite{Sokhotski,*Plemelj}, infinitesimally close to the real axis, the imaginary part of the self-energies can be determined by a simple one-dimensional integration over $\phi$ 
\be
{\rm Im}  [ \Sigma (z) ]= - m_q^2  \ \Theta(1-z^2) \int_0^{2\pi}  \frac{d\phi}{4 k} \ \frac{\gamma_0 + {\bf \hat{v} \cdot \boldsymbol \gamma}}{\left[H(z, \phi)\right]^2} \, , \label{qse3}
\ee
for quarks, and
\be
{\rm Im}  [ \Pi^{ij}(z) ]= - m_D^2 \ \Theta(1-z^2) \int_0^{2\pi} \frac{d \phi}{4 k}  \ \frac{v^i  k^l v^j}{ \left[H_\Xi(z,\phi)\right]^4}\ W^l \, , \label{gse3}
\ee
for gluons, where $\Theta(x)$ is the unit step (Heaviside) function.

The integrals \eqref{qse3} and \eqref{gse3} can be calculated numerically without any complications, because the integrands are non-singular and the integration region is bounded.  In Appendix \ref{appndxa}, we also provide a method to obtain analytic results for these kinds of integrations for a general class of deformations $H({\bf\hat{v}};\{ \boldsymbol \alpha \})$, and there we give an example result for the case of spheroidal anisotropy. However, in general, the analytic results can possess a large number of terms which makes their evaluation time-consuming and impractical to use. In plasma coordinates (primed), it may be possible to obtain simpler analytic forms for the first step integrations over $\phi'$ as was done in our previous paper \cite{Kasmaei:2016apv}, however, the second integration over $x'$ variable remains non-trivial, with multiple parametric poles and branch cuts. 

The real part of the self-energy integrals can be calculated directly as the numerical principal value of the singular integrands with first-order poles. However, due to the singularity and the necessity of obtaining the result in a large range of $z$, this procedure is not efficient enough for practical use. To proceed, we introduce an efficient method to determine the real parts of self-energies based solely on the numerical calculation of the imaginary parts from Eqs.~\eqref{qse3} and \eqref{gse3}. This alternative method lacks the complications related to numerical integration of singular functions, and allows one to obtain results in times which are orders of magnitude less than the direct numerical integration and with no loss of accuracy. This method provides, not only numerical values, but also analytic formulae for the real parts of self-energy applicable to all values of the complex phase velocity $z$. This method is based on the fact that the real part of the self-energy, e.g. quark self-energy, can be written as
\be
{\rm Re}  [ \Sigma (z) ] = - \int_{-1}^1  \frac{dx}{\pi} \ \frac{{\rm Im} [ \Sigma (x) ] }{z-x + i0^+} \, .
\ee
Based on this expression, if an accurate approximation of ${\rm Im} [ \Sigma^{\mu} (x) ]$ in the interval \mbox{$x \in [-1,1]$} can be found as a series in terms of some basis functions $r_n(x)$, i.e.
\be
{\rm Im} [ \Sigma^{\mu} (x) ] \approx  \sum_{n=0}^{n_{\rm max}} c_n^{\mu} \,  r_n (x) \, ,
\ee
and provided that the integral $I_n(z)= \int_{-1}^1 dx \, \frac{r_n(x)}{z-x+i 0^+}$ has a simple analytic form, the real part can be written as
\be
{\rm Re} [ \Sigma^{\mu} (z) ] \approx  -\frac{1}{\pi}\sum_{n=0}^{n_{\rm max}} c_n^{\mu} \, {\rm Re} [ I_n (z) ] \, .
\ee
In particular, if a finite polynomial series can well-approximate the imaginary part, setting $r_n(x)= x^n$, one has
\be
I_n(z) = \! \int_{-1}^1 dx  \frac{x^n}{z-x+i 0^+} =  \frac{  _{2}F_{1}(1,1+n,2+n,-1/z) \ +  (-1)^{n} \, _{2}F_{1}(1,1+n,2+n, 1/z)}{(1+n) z} \, , 
\ee
where $_{2}F_{1}$ is the hypergeometric function. In Appendix \ref{appndxb}, we present recursive relations for the $I_n(z)$ functions, their large $z$ behavior, and some useful relations obeyed by the underlying hypergeometric functions.  Note that the basis functions $I_n(z)$ are valid for the calculation of the self-energies at all points of the complex plane.

The approximating series, the basis functions, the maximum order, and the coefficients of the expansion are not unique. Indeed, the approximation of the imaginary part can be piecewise or even pointwise.  In the case of a piecewise approximation, the interval $x \in [-1,1]$ is divided to sub-intervals  $[ x_i,x_{i+1}]$, and for each sub-interval a set of coefficients $\{c_n^{\mu,(i)}\}$ need to be found to approximate the values of ${\rm Im} [ \Sigma^{\mu} (x) ]$ for $x \in [ x_i,x_{i+1} ]$ using the basis functions $ r_n (x)$. Then, with $x_0=-1$ and $x_{m+1}=1$,  
\be
{\rm Re} [ \Sigma^{\mu} (z) ] \approx  -\frac{1}{\pi} \sum_{i=0}^m \sum_{n=0}^{n^{(i)}_{\rm max}} c_n^{\mu,(i)} \, {\rm Re} [ I^{(i)}_n (z) ] \, .
\ee
In the case of polynomial/hypergeometric basis functions for the piecewise approximation of imaginary/real parts of self-energy integrals, one can use
\ba
I^{(i)}_n(z) \ &=& \int_{x_i}^{x_{i+1}} dx  \frac{x^n}{z-x+i 0^+} \nonumber \\ 
&=& \ \frac{  x_{i}  \ _{2}F_{1}(1,1+n,2+n,x_i /z) \ -  (x_{i+1})^{n+1} \ _{2}F_{1}(1,1+n,2+n, x_{i+1}/z)}{(1+n) z} \, . 
\ea
Using a piecewise approximation can in general reduce the maximum order of the basis functions, but since the total number of coefficients increases with the number of sub-intervals, there is a trade-off between the number of sub-intervals and the maximum order. In the particular case of ellipsoidal anisotropy, we obtained desirable accuracy and speed of computation by fitting only in the single interval $[-1,1]$. On the other extreme side, one may use a pointwise approximation of the imaginary part, which requires only the zeroth-order basis functions. However, in the pointwise method the number of terms in the approximating function will be large, and the computational speed will be reduced to the case of a direct numerical integration using a typical quadrature method.

Summarizing, for a given component of the quark self-energy, the standard hypergeometric expansion method consists of evaluating the imaginary part of said self-energy component in the real space-like region $z \in [-1,1]$ where the integrals are straightforward to evaluate and then numerically fitting the result to extract a finite set of polynomial coefficients $c_n^\mu$.  Once these coefficients are extracted, the result for both the real and imaginary parts can be well-approximated in the entire complex plane using the hypergeometric basis function expansion.  In practice, we find that 20 coefficients provide results with a relative error of less than $10^{-6}$ in the entire complex plane even for extremely anisotropic distributions.  For weakly anisotropic distributions, even fewer coefficients are required to achieve the same accuracy.

In the same way, the real part of the gluon self-energy can be calculated as
\ba
{\rm Re} [ \Pi^{ij} (z) ] &=&  \Pi_0^{ij} +  \Pi_k^{ij} (z)   \\
&=&   m_D^2 \int \frac{d \Omega} {4\pi} \ v^i \ \frac{W^l(\theta,\phi)}{ \left[H_\Xi(\theta,\phi)\right]^4} \ \delta^{jl} \   - \int_{-1}^1  \frac{dx}{\pi} \ \frac{{\rm Im} [  \Pi^{ij} (x) ] }{z-x + i0^+}    \nonumber \\
  &\approx & m_D^2 \int \frac{d \Omega}{4\pi} \  v^i \ \frac{W^j(\theta,\phi)}{ \left[H_\Xi(\theta,\phi)\right]^4} \  -\frac{1}{\pi}\sum_{n=0}^{n_{\rm max}} c^{ij}_n \ {\rm Re} [ I_n (z) ] \, , \nonumber
\ea
with the $c^{ij}_n$ coefficients being calculated by fitting the expansion of the imaginary part of the gluon self-energy as
\be
{\rm Im} [  \Pi^{ij} (x) ] \approx  \sum_{n=0}^{n_{\rm max}} c^{ij}_n \,  r_n (x) \, .
\ee
\section{Results}
\label{results}

In this section we calculate the quark and gluon self-energies for a QGP with an ellipsoidal momentum anisotropy in several cases, including special cases of an isotropic and spheroidally anisotropic QGP. We also present plots of the dispersion relations $\omega(\bf k)$ for the quark and gluon collective modes and present the growth rate $\gamma(\bf k)$ for unstable gluon modes for some specific values of the anisotropy parameters and propagation direction. Note that all results presented in this section are scaled by dividing by the quark or gluon effective mass scales $m^2_q$ and $m^2_D$ defined in Eqs.~\eqref{mq2} and \eqref{md2}.


\subsection{Quark self-energies and collective modes}
\label{quark}

The quark self-energy in an ellipsoidally-anisotropic QGP is
\be
\Sigma(z;\{ \xi_1, \xi_2, {\bf \hat{n}_1}, {\bf \hat{n}_2} \}) = \frac{m_q^2}{4\pi k} \int_{-1}^1 dx \int_0^{2\pi} d\phi\ \frac{1}{1+ \xi_1 ({\bf \hat{n}_1 \cdot \hat{v}})^2 +  \xi_2 ({\bf \hat{n}_2 \cdot \hat{v}})^2 } \  \frac{\gamma_0 + {\bf \hat{v} \cdot \boldsymbol \gamma}}{z - x + i0^{+} } \, . 
\ee
With the choice of parton coordinates as defined in Sec.~\ref{frame}, the ellipsoidal anisotropy function with parameters $\{\theta_k, \phi_k, \xi_1, \xi_2\}$ contains
\be
\begin{split}
&\xi_1 ({\bf \hat{n}_1 \cdot \hat{v}})^2 +  \xi_2 ({\bf \hat{n}_2 \cdot \hat{v}})^2 =\  \xi_1 \big (\cos \theta_k \ x - \sin \theta_k \ \sin \phi \ \sqrt{1-x^2} \big )^2 \\
& + \ \xi_2 \big (\sin \theta_k \cos \phi_k \ x + \cos \theta_k \cos \phi_k \ \sqrt{1-x^2} \sin \phi + \sin \phi_k \ \sqrt{1-x^2} \cos \phi \big )^2 \, .
\end{split}
\ee
For the special case of an isotropic QGP, $\xi_1=\xi_2=0$, the analytic result for real $z$ is
\ba
 \Sigma^0 (z,{\bf k}) &=&  \frac{m_q^2}{2 k} \left[  \log \left|\frac{z+1}{z-1}\right|- i\pi \Theta(1-z^2) \ \right] , \\
 \boldsymbol \Sigma (z,{\bf k}) &=& \frac{m_q^2}{2 k} \left[-2 + z \log \left|\frac{z+1}{z-1}\right|- i\pi z \Theta(1-z^2) \ \right] \bf\hat{k} \, .
\ea

For a momentum-anisotropic QGP, it is possible to obtain analytic results for the imaginary parts of the parton self-energies by integrating over $\phi$ using the residue theorem or the fraction splitting method.   In App.~\ref{appndxa}, we present analytic results for the imaginary part of the quark self-energy as an example. The analytic result for the case of an ellipsoidal anisotropy has a very large number of terms and their evaluation becomes time-consuming.  The integration method using the hypergeometric expansion proposed in Sec.~\ref{methods} provides fast and accurate results for the integrals of self-energies for all types and strengths of anisotropy, valid for all complex values of $(\omega, {\bf k})$.  In all results reported in the remainder of the paper, we used the hypergeometric expansion method.

\begin{figure}[H]
\centerline{
\includegraphics[width=0.95\linewidth]{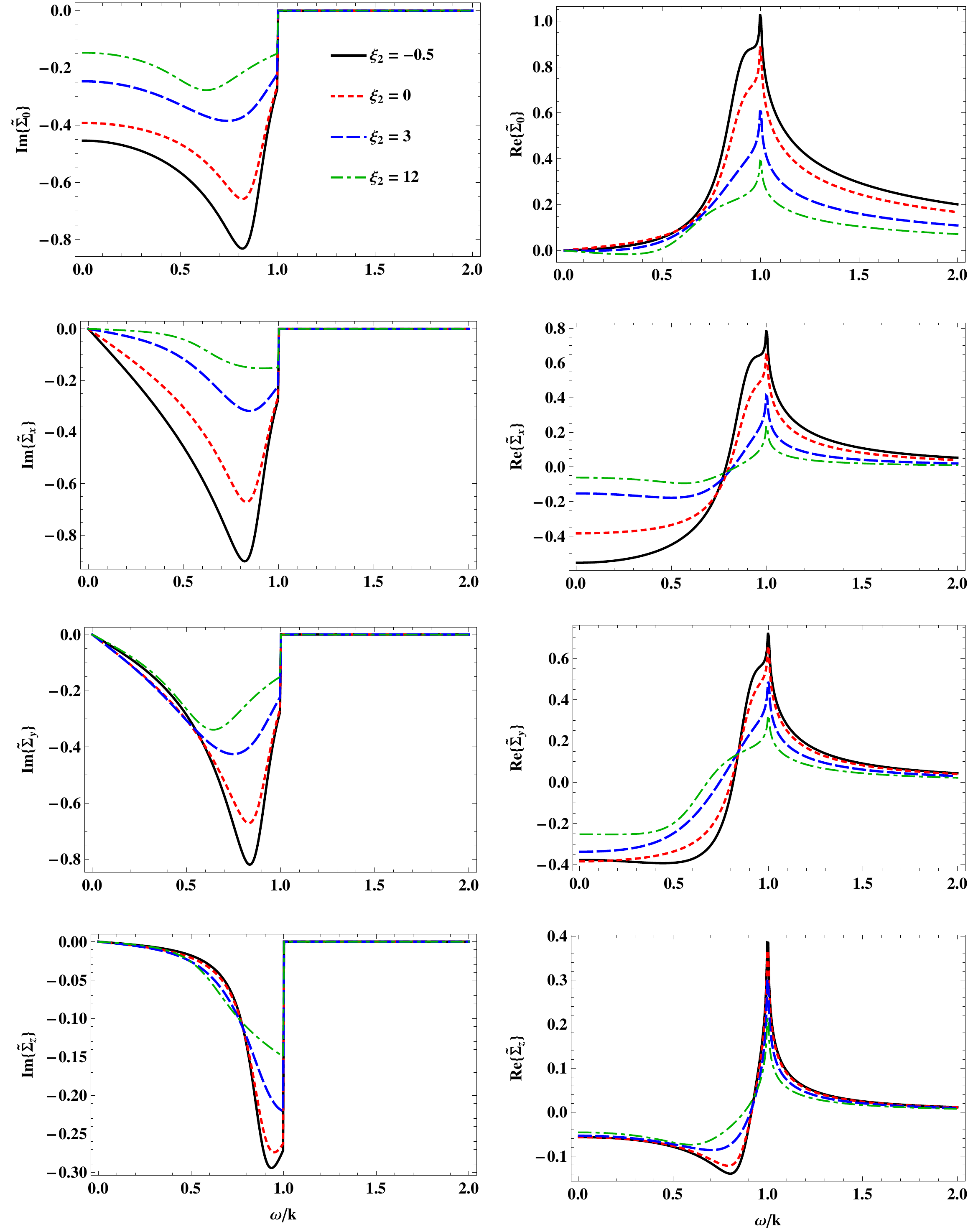}
}
\caption{The real and imaginary parts of $\tilde{\Sigma}_0$, $\tilde{\Sigma}_x$, $\tilde{\Sigma}_y$, and $\tilde{\Sigma}_z$ as a function of $\omega/k$ for $\xi_1=20$, $\theta_k=\pi/3$, $\phi_k=\pi/4$, and $\xi_2=\{-0.5,0,3,12\}$.} 
\label{plot:plot1}
\end{figure}

\begin{figure}[H]
\centerline{
\includegraphics[width=0.95\linewidth]{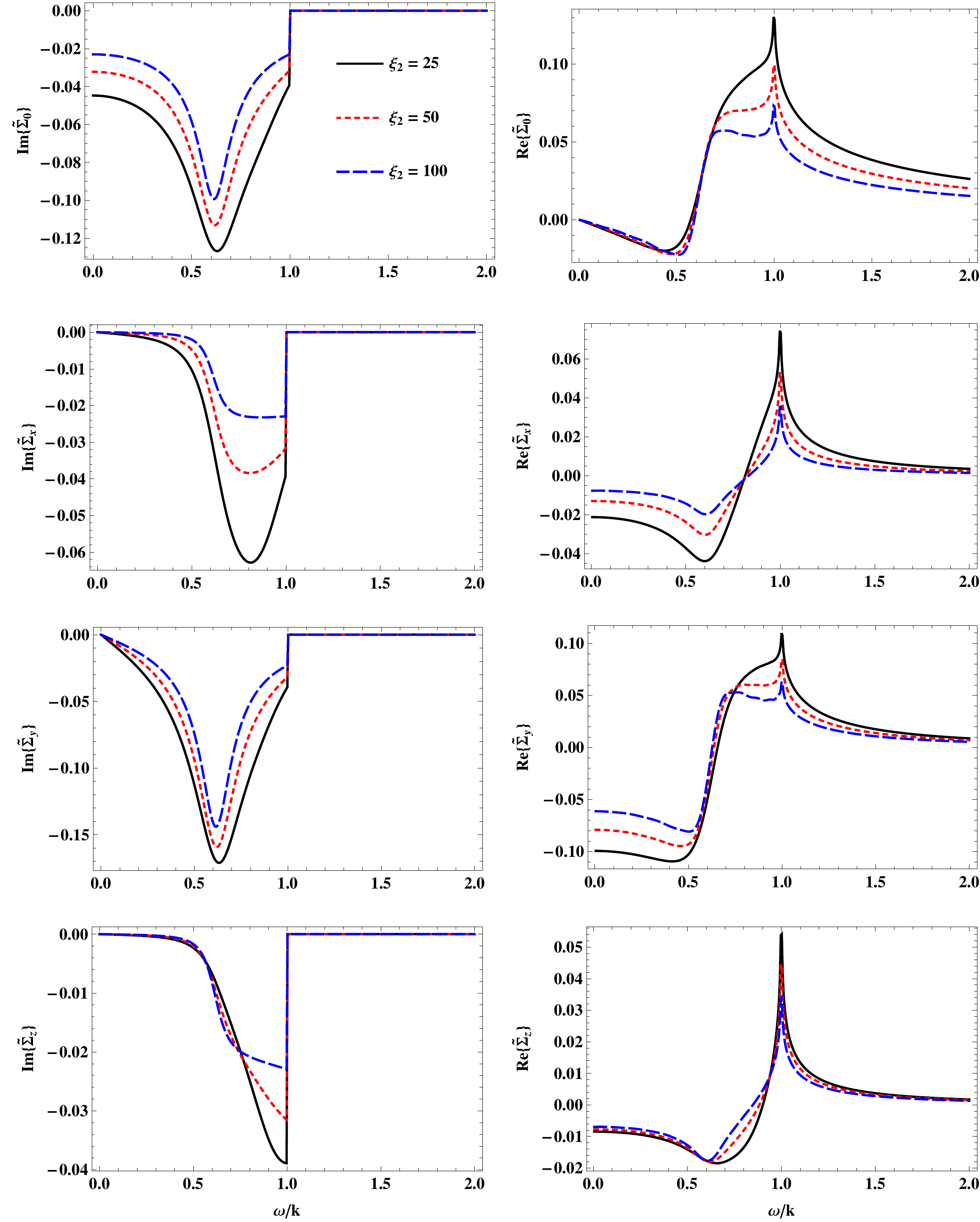}
}
\caption{The real and imaginary parts of $\tilde{\Sigma}_0$, $\tilde{\Sigma}_x$, $\tilde{\Sigma}_y$, and $\tilde{\Sigma}_z$ as a function of $\omega/k$ for $\xi_1=120$, $\theta_k=\pi/3$, $\phi_k=\pi/4$, and $\xi_2=\{25,50,100\}$.} 
\label{plot:plot2}
\end{figure}

\begin{figure}[H]
\centerline{
\includegraphics[width=0.95\linewidth]{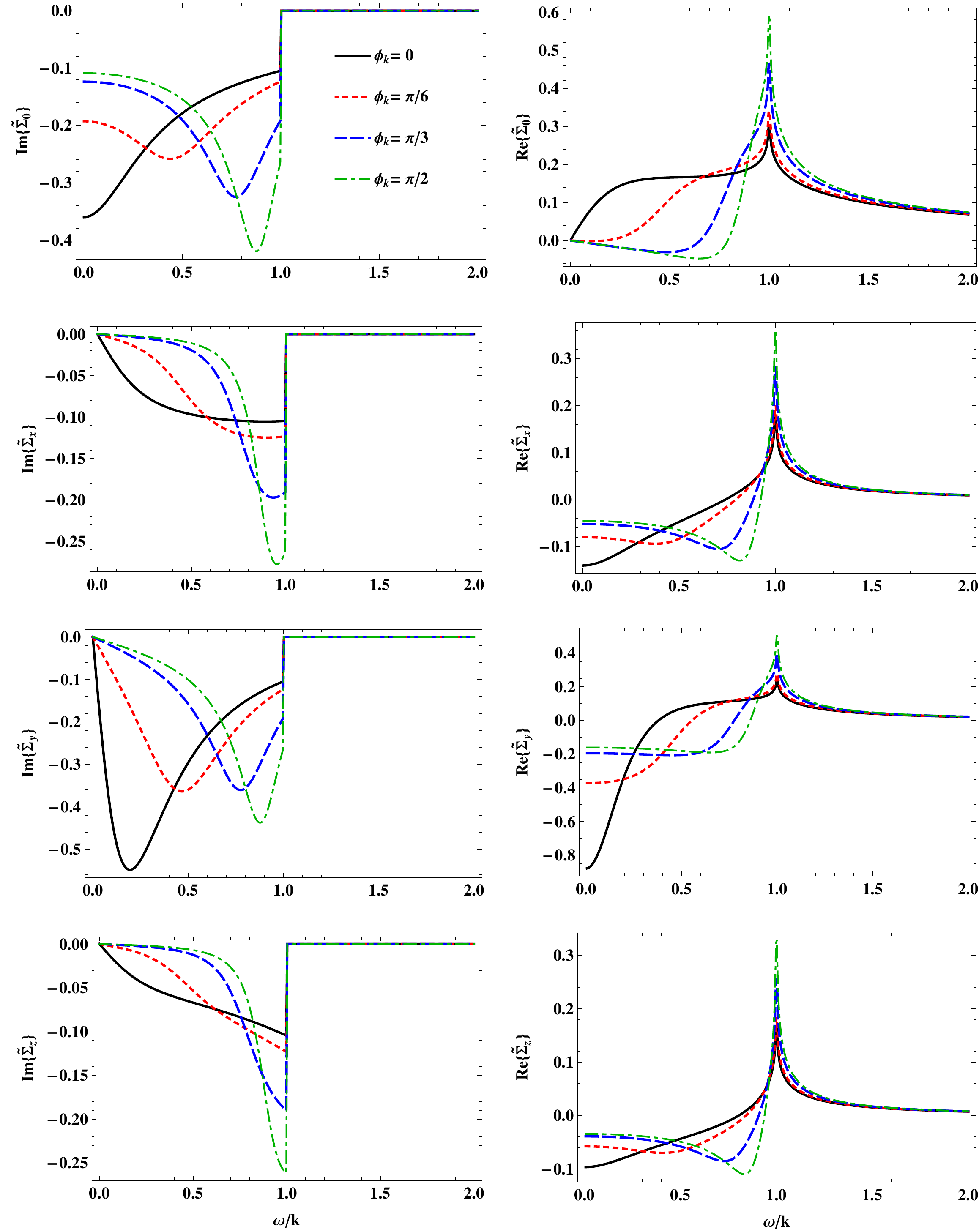}
}
\caption{The real and imaginary parts of $\tilde{\Sigma}_0$, $\tilde{\Sigma}_x$, $\tilde{\Sigma}_y$, and $\tilde{\Sigma}_z$ as a function of $\omega/k$ for $\xi_1=20$, $\theta_k=\pi/3$, $\xi_2=12$, and $\phi_k=\{0,\pi/6,\pi/3,\pi/2\}$.} 
\label{plot:plot3}
\end{figure}

In Fig.~\ref{plot:plot1}, we plot the real and imaginary parts of \mbox{$\tilde{\Sigma}^{\mu} (z) \equiv k^2\Sigma(z,{\bf k})/(k_{\mu} m_q^2)$} in the plasma coordinate system where \mbox{$k^{\mu}= k (1, \sin\theta_k \cos\phi_k, \sin\theta_k \sin\phi_k, \cos\theta_k)$} is the quark four-momentum.  The left and right columns show the imaginary and real parts, respectively.  The four lines in each panel show different values of the azimuthal anisotropy strength parameter $\xi_2$ in the ellipsoidal parameterization, with the values of $\xi_1, \theta_k, \phi_k$ held fixed.  The rows from top to bottom show $\tilde{\Sigma}^0$, $\tilde{\Sigma}^x$, $\tilde{\Sigma}^y$, and $\tilde{\Sigma}^z$, respectively.  As can be seen from these plots, the method yields high accuracy results even in the vicinity of $z=1$ where there is a logarithmic cusp.  As further evidence of the reliability of the hypergeometric expansion method, in Fig.~\ref{plot:plot2} the quark self-energy components are shown for a highly anisotropic plasma where $\xi_1=120$ and $\xi_2 \in \{25, 50, 100\}$. The results with several values of $\phi_k \in \{0,\pi/6,\pi/3,\pi/2\}$, with other parameters held fixed, are shown in Fig.~\ref{plot:plot3}.  Figs.~\ref{plot:plot2} illustrates the reliability of the method even in the case of extreme momentum anisotropy and Fig.~\ref{plot:plot3} demonstrates the explicit dependence of the quark self-energy on the azimuthal angle when $\xi_2 \neq 0$.

\clearpage

\begin{figure}[H]
\vspace{5mm}
\centerline{
\includegraphics[width=0.8\linewidth]{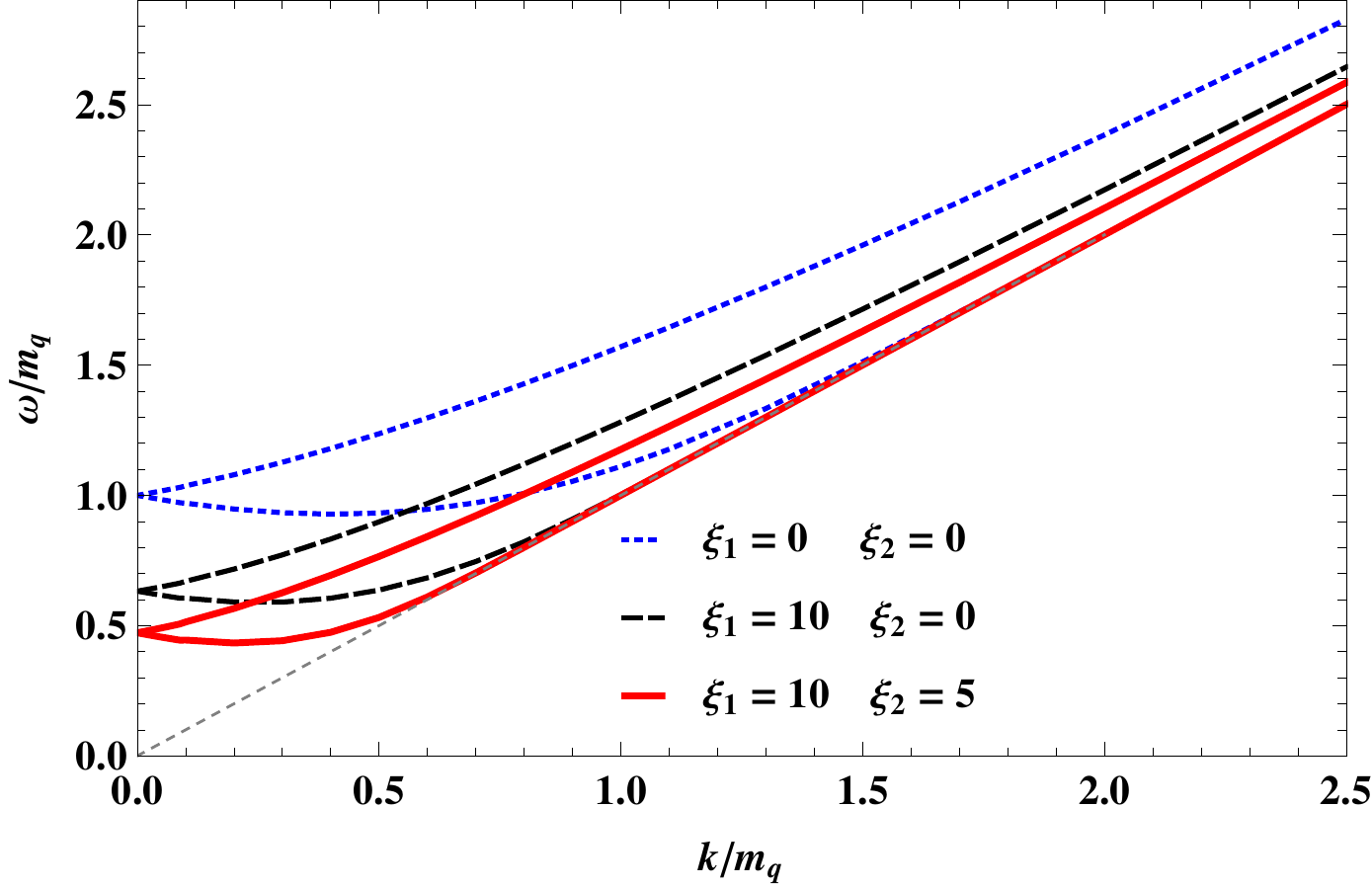}
}
\caption{Dispersion relations for the quark collective modes in isotropic ($\xi_1=\xi_2=0$), spheroidally-anisotropic ($\xi_1=10, \ \xi_2=0$), and ellipsoidally-anisotropic ($\xi_1=10, \ \xi_2=5$) mediums, for $\theta_k=\pi/3$ and $\phi_k=\pi/6$.} 
\label{plot:clctvq}
\end{figure}

The dispersion relations for the quark collective modes in a QGP are given by the roots of the characteristic equation
\be 
\big[\omega - \Sigma^0(\omega,{\bf k})\big]^2- \big[{\bf k} - \boldsymbol \Sigma(\omega,{\bf k})\big]^2 = 0 \, .  \label{chrceqq}
\ee 
Solving Eq.~\eqref{chrceqq} for $\omega(\bf k)$, the dispersion relations for the quark collective modes for isotropic, spheroidally-anisotropic, and ellipsoidally-anisotropic momentum distributions are plotted in Fig.~\ref{plot:clctvq}. For each of these three cases, there are only two stable (real $\omega$) and no unstable quark collective modes (imaginary $\omega=i\gamma$).  This agrees with the findings of Ref.~\cite{Schenke:2006}.  However, the dispersion relations for the collective modes in an anisotropic QGP naturally depend on the direction of the quark momentum which includes azimuthal angle dependence in the case of an ellipsoidally-anisotropic QGP.

\subsection{Gluon self-energies and collective modes}
\label{gluon}

The components of the gluon self-energy in an ellipsoidally-anisotropic QGP can be obtained from
\be
\Pi^{ij}(\omega, {\bf k})= m_D^2 \int \frac{d \Omega}{4\pi} v^i \ \frac{v^l + \xi_1({\bf n_1 \cdot v})n^l + \xi_2({\bf n_2 \cdot v})n^l  }{ \big( \ 1+ \xi_1 ({\bf n_1 \cdot v})^2 +  \xi_2 ({\bf n_2 \cdot v})^2 \ \big)^2} \  \left[\delta^{j l} + \frac{k^l v^j}{\omega -{\bf k \cdot v}  +i 0^{+}} \right] . \label{gse4}
\ee
Since $\Pi^{ij}$ is a symmetric three-tensor, there are six independent components of $\Pi^{ij}(\omega, {\bf k})$.

\begin{figure}[H]
\centerline{
\includegraphics[width=0.95\linewidth]{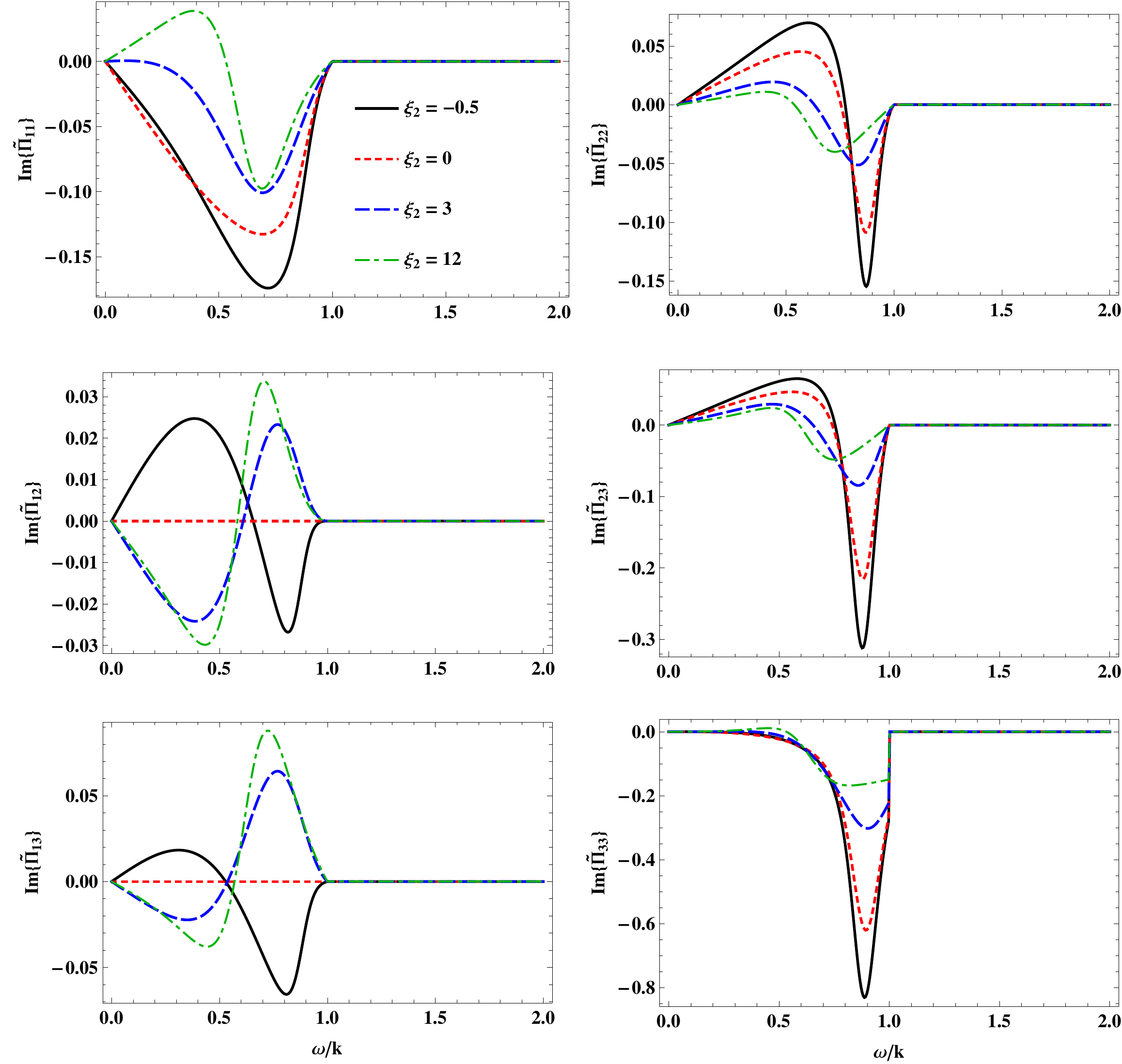}
}
\caption{The imaginary parts of $\tilde{\Pi}_{11}$, $\tilde{\Pi}_{12}$, $\tilde{\Pi}_{13}$, $\tilde{\Pi}_{22}$, $\tilde{\Pi}_{23}$, and $\tilde{\Pi}_{33}$ as a function of $\omega/k$ for $\xi_1=20$, $\theta_k=\pi/3$, $\phi_k=\pi/4$, and $\xi_2=\{-0.5,0,3,12\}$.} 
\label{plot:plot4}
\end{figure}

\begin{figure}[H]
\centerline{
\includegraphics[width=0.95\linewidth]{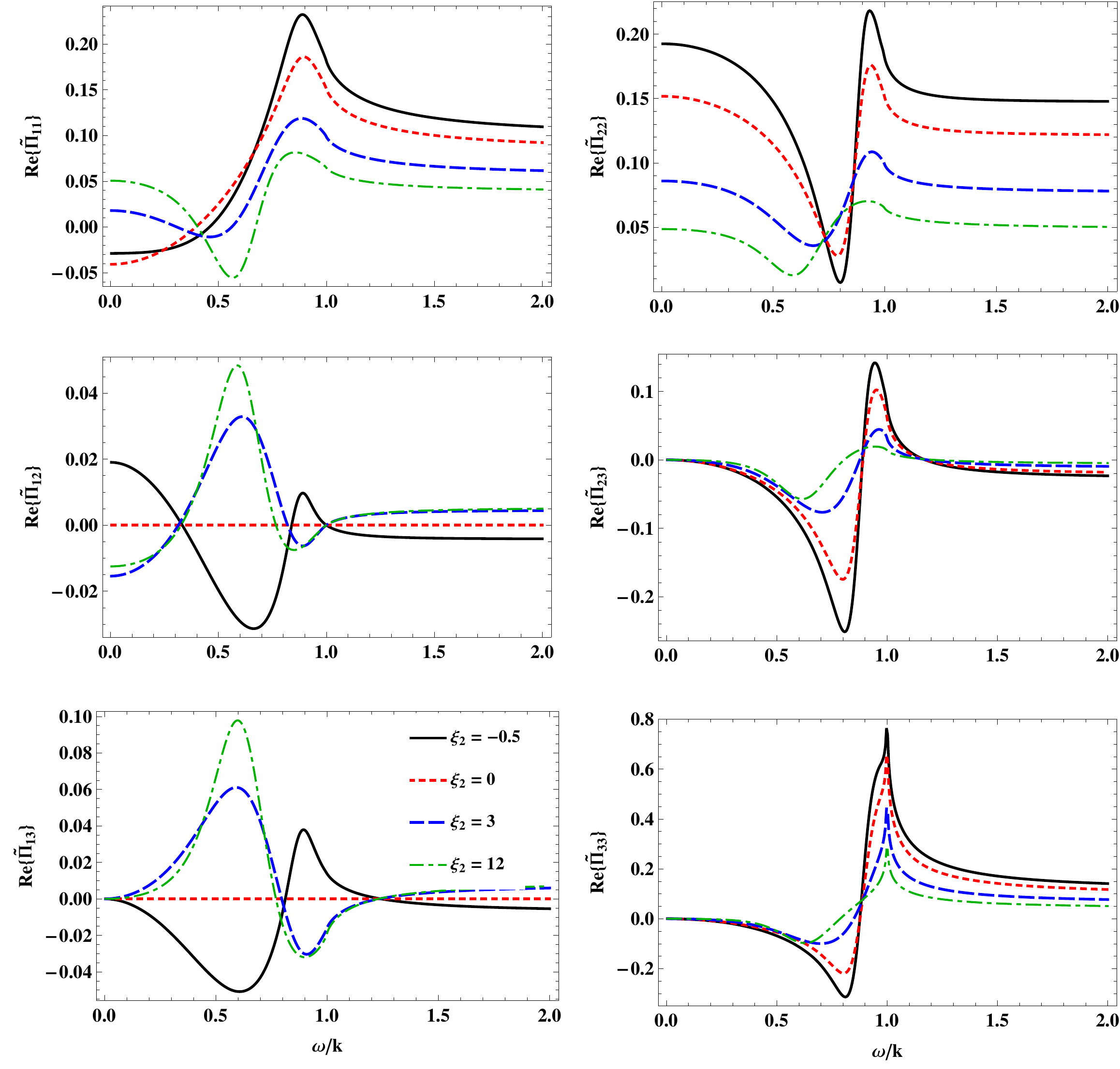}
}
\caption{The real parts of $\tilde{\Pi}_{11}$, $\tilde{\Pi}_{12}$, $\tilde{\Pi}_{13}$, $\tilde{\Pi}_{22}$, $\tilde{\Pi}_{23}$, and $\tilde{\Pi}_{33}$ as a function of $\omega/k$ for $\xi_1=20$, $\theta_k=\pi/3$, $\phi_k=\pi/4$, and $\xi_2=\{-0.5,0,3,12\}$.} 
\label{plot:plot5}
\end{figure}

\begin{figure}[H]
\centerline{
\includegraphics[width=0.95\linewidth]{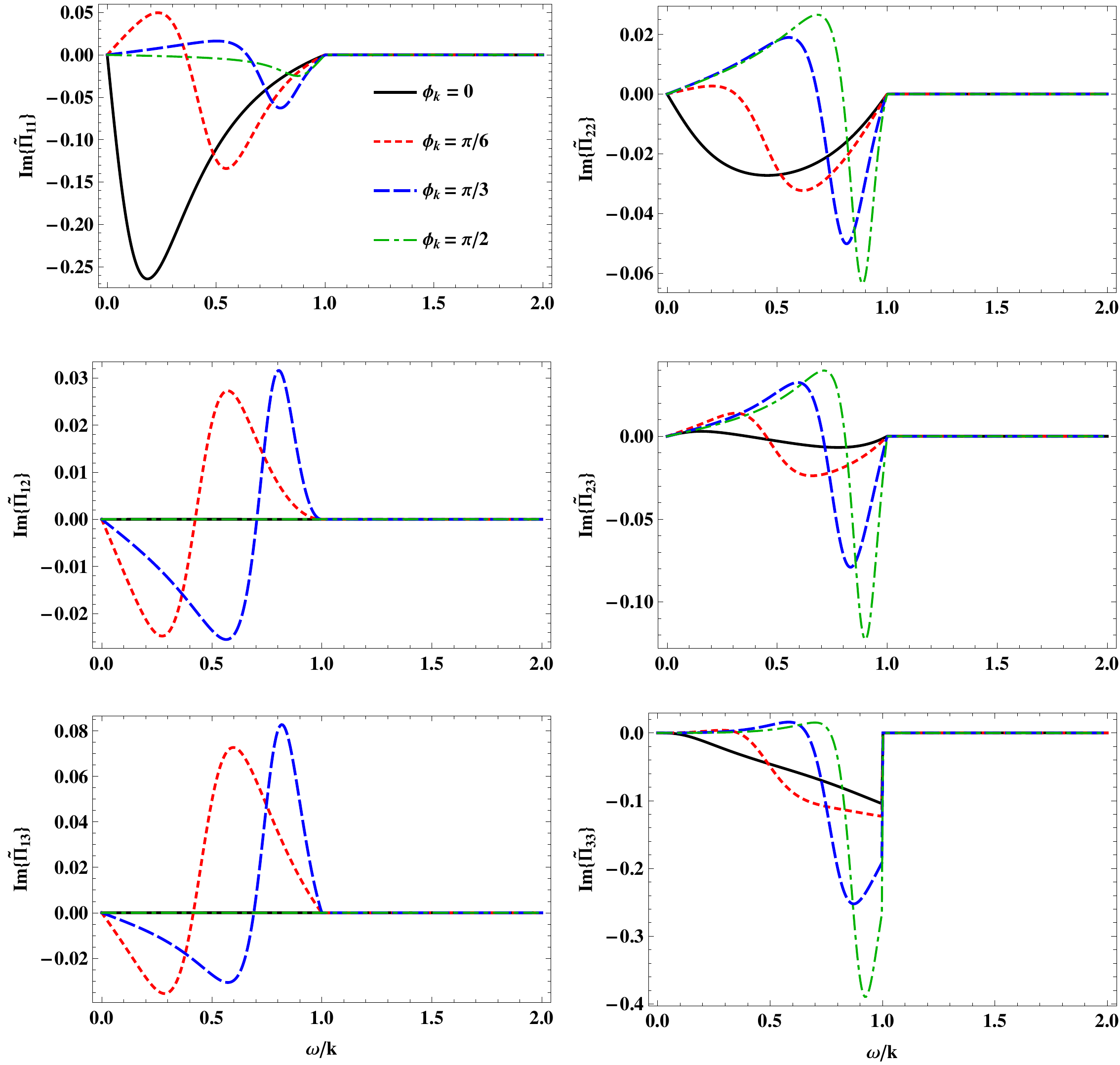}
}
\caption{The imaginary parts of $\tilde{\Pi}_{11}$, $\tilde{\Pi}_{12}$, $\tilde{\Pi}_{13}$, $\tilde{\Pi}_{22}$, $\tilde{\Pi}_{23}$, and $\tilde{\Pi}_{33}$ as a function of $\omega/k$ for $\xi_1=20$, $\theta_k=\pi/3$, $\xi_2=12$, and $\phi_k=\{0,\pi/6, \pi/3, \pi/2\}$.} 
\label{plot:plot6}
\end{figure}

\begin{figure}[H]
\centerline{
\includegraphics[width=0.95\linewidth]{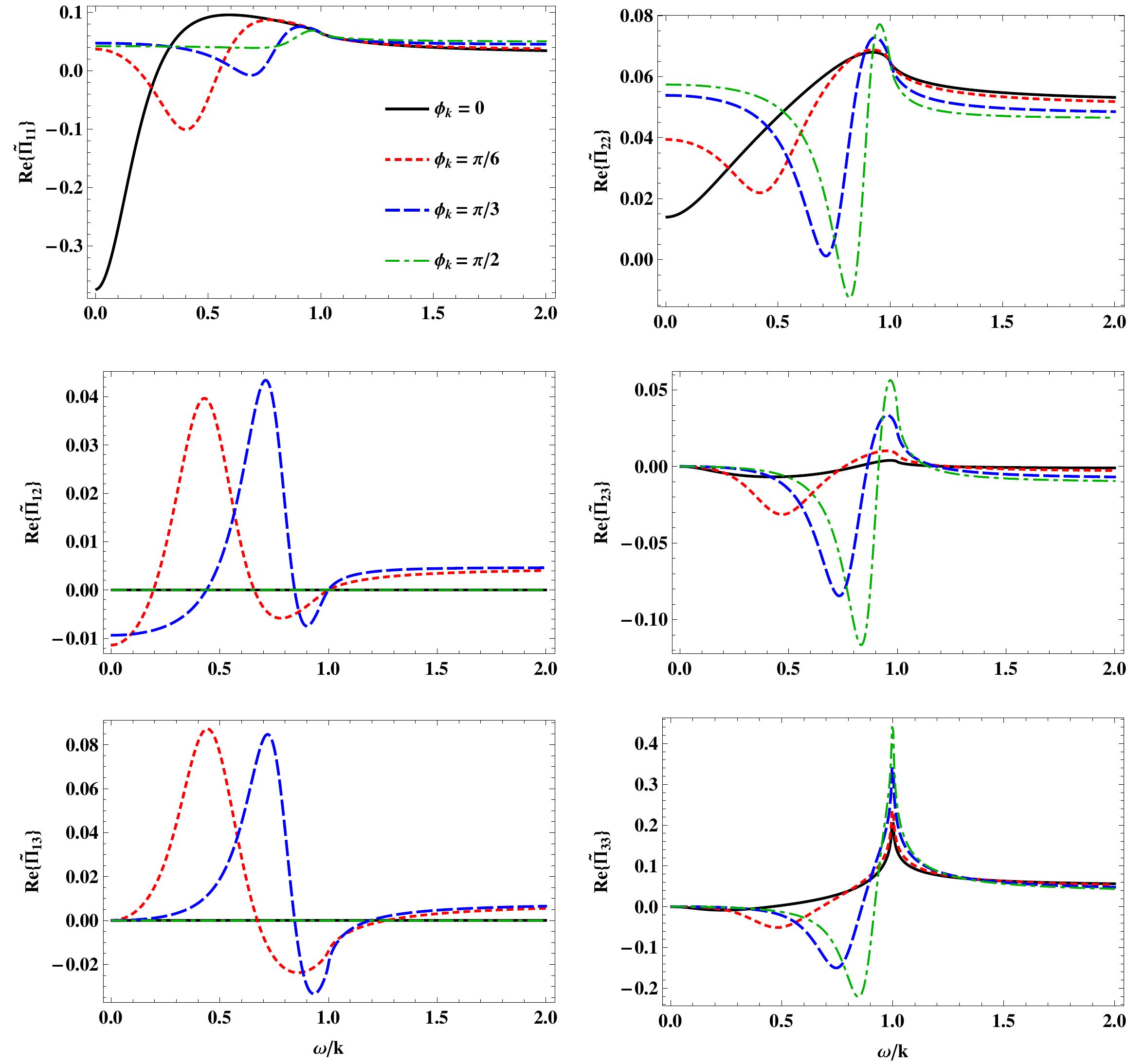}
}
\caption{The real parts of $\tilde{\Pi}_{11}$, $\tilde{\Pi}_{12}$, $\tilde{\Pi}_{13}$, $\tilde{\Pi}_{22}$, $\tilde{\Pi}_{23}$, and $\tilde{\Pi}_{33}$ as a function of $\omega/k$ for $\xi_1=20$, $\theta_k=\pi/3$, $\xi_2=12$, and $\phi_k=\{0,\pi/6, \pi/3, \pi/2\}$.} 
\label{plot:plot7}
\end{figure}

In Fig.~\ref{plot:plot4} we plot the imaginary parts of independent components of $\tilde\Pi^{ij} \equiv \, \Pi^{ij}/m_D^2$ for different values of the second anisotropy strength parameter $\xi_2$ of ellipsoidal parameterization with the values of $\xi_1, \theta_k, \phi_k$ held fixed. The corresponding real parts are shown in Fig.~\ref{plot:plot5}. The results of variation of $\phi_k$, with other parameters held fixed, are shown in Figs~\ref{plot:plot6} and \ref{plot:plot7}. 

The gluon collective modes can be obtained from the roots of the characteristic equation 
\be
\det[ ({\bf k}^2 - \omega^2)\delta^{ij} - k^i k^j - \Pi^{ij} ] = 0 \, ,
\ee
With the choice of the parton coordinate system where ${\bf k} = (0,0,k)$, this simplifies to
\ba
&&\Pi _{13} \left[(\omega^2-k^2) \Pi _{13}+\Pi _{22}\Pi _{13} -\Pi _{12} \Pi
   _{23}\right] + \Pi _{23} \left[ (\omega^2-k^2)\Pi _{23}
   +\Pi _{11} \Pi_{23} -\Pi _{13} \Pi _{21}\right] \nonumber \\ 
&&\hspace{3.5cm}
 - \left(\Pi _{33}+\omega^2\right) \left[\left(k^2-\omega^2-\Pi _{11}\right) \left(k^2-\omega^2-\Pi _{22}\right)-\Pi _{12}^2\right] = 0 \, .  \hspace{5mm} \label{chrceqg}
\ea

\begin{figure}[H]
\centerline{
\includegraphics[width=0.95\linewidth]{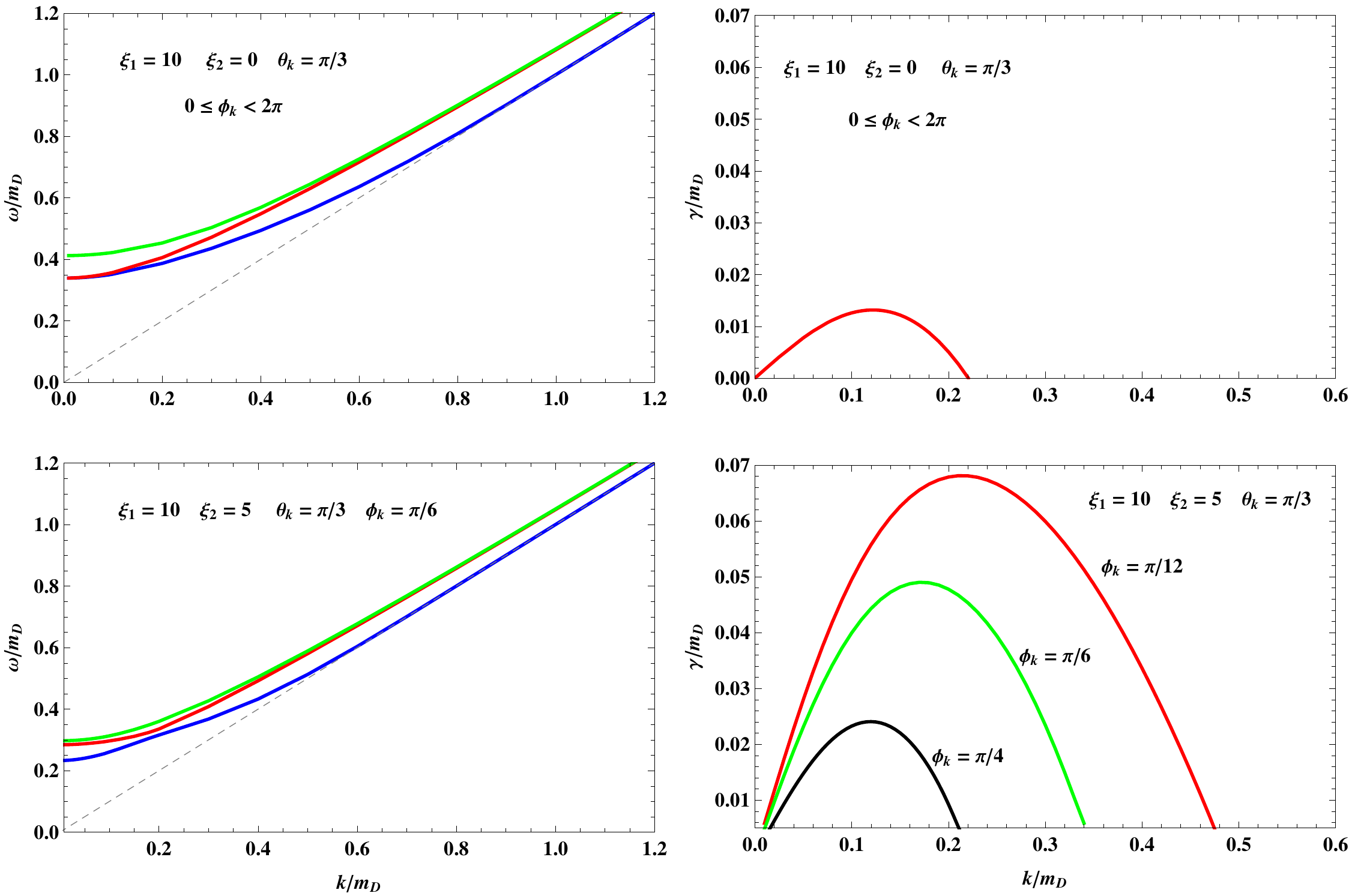}
}
\caption{Dispersion relations for gluon collective modes for mediums with spheroidal ($\xi_1=10,\ \xi_2=0$) and ellipsoidal ($\xi_1=10,\ \xi_2=5$) momentum anisotropy. The unstable modes (imaginary $\omega=i \gamma$) in the ellipsoidally anisotropic case depend on $\phi_k$.} 
\label{plot:clctvg}
\end{figure}

The dispersion relations for the gluon collective modes, obtained by numerically solving Eq.~\eqref{chrceqg}, are shown in Fig.~\ref{plot:clctvg} for two cases with spheroidal and ellipsoidal momentum anisotropy. We find that there are three stable modes (with real $\omega$) in each case. The number of unstable modes (with imaginary $\omega=i \gamma$) depend on the direction of the gluon momentum. With the introduction of the second parameter of anisotropy $\xi_2$, the unstable modes growth rates depend strongly on $\phi_k$, and for some directions they can grow several times larger than in the case of the spheroidal anisotropy.  

\section{Conclusions and outlook}
\label{conclusion}

In this paper we discussed collective excitations in a momentum-anisotropic quark-gluon plasma. We introduced an efficient method for calculating both the quark and gluon self-energies which is applicable to most types of anisotropic distribution functions, to all ranges of the strength of such anisotropies, and to all values of momentum and complex energy. This method exploits the smoothness and boundedness of the imaginary parts of the space-like self-energies and, with a proper choice of the coordinate system, combines a fit-based series expansion of the imaginary part results with analytically available integrals in order to extract the results for the singular integrals of the real parts of the self-energies as a weighted sum of a class of basis functions e.g. hypergeometric functions. Because the integrands of the imaginary parts are smooth and non-singular, their direct numerical calculation is fast and straightforward. When available, one can also use analytic results for this first step of the calculation. A method to extract analytic results for the imaginary parts of self-energies in anisotropic mediums, with a few examples, was also introduced in an appendix to this paper. Using the fitting method, the imaginary part can be approximated by a finite-order polynomial in the phase velocity. The coefficients resulting from such an expansion, when multiplied by the appropriate hypergeometric basis functions, allow one to compute both the real and imaginary parts of the parton self-energies in the entire complex phase-velocity plane. 

To demonstrate the application of the proposed method, and to also extend the results of previous works, several results for the parton self-energies and dispersion relations in a QGP with ellipsoidal momentum anisotropy were provided in this paper.  The results possess the full direction dependence of collective excitations in an ellipsoidally-anisotropic QGP characterized by two anisotropy directions and strengths. These results are relevant to the phenomenological studies of various physical aspects of QGP.  In particular, the space-like quark self-energy enters the calculation of real and virtual photon production rates from a QGP in the hard-loop framework~\cite{Schenke:2007, Bhattacharya:2016, Baym:2017qxy}. The unstable gluon modes play an important role in the thermalization of a weakly-coupled QGP~\cite{Mrowczynski:2016etf}. The gluon self-energies are also relevant to the studies of bound states in a hot QGP~\cite{Dumitru:2007hy,Burnier:2009yu,Dumitru:2009fy,Nopoush:2017zbu}, the heavy quark potential, and quarkonia suppressions \cite{Strickland:2011mw,Strickland:2011aa,Krouppa:2016jcl,Krouppa:2017lsw}.  The method introduced in this paper, provides an efficient way to extract parton self-energies to be used in such secondary calculations of physical observables where, in most cases, an extremely large number of self-energy evaluations becomes necessary.  Comparing to conventional numerical integration methods, the method introduced here was always more than 1000 times faster, with no loss of accuracy.  The computational gains can be further enhanced by tabulating the coefficients determined by fits to the imaginary part of the various self-energies and interpolating them in momentum-deformation parameter space.  Using such coefficient interpolations, one can compute the necessary self-energies  in the entire complex plane approximately $10^{9}$ times faster than direct numerical integration.

In future works, various phenomenological aspects of QGP with two independent anisotropy directions and strength parameters can be studied, including photon and dilepton production, color plasma instabilities, and bound states in the QGP.  The results of this paper can be extended in a similar way to even more general anisotropic momentum distribution functions, e.g. by including off-diagonal terms in the quadratic anisotropy parameterization \eqref{quadratic}.

\acknowledgments

M.\ Strickland and B.\ Kasmaei were supported by the U.S. Department of Energy, Office of Science, Office of Nuclear Physics under Award No.~DE-SC0013470.

\appendix 

\section{Analytic results for imaginary parts of self-energies}
\label{appndxa}

Using a fraction splitting method, one can obtain analytic results for the imaginary parts of the self-energy integrals. In the cases where the integrals \eqref{qse3} and \eqref{gse3} over $\phi$ can be written as 
\be
\int_0^{2\pi} d\phi \  \frac{\sum_m C_m \sin^m \phi + D_m \cos^m \phi }{\sum_n A_n \sin^n \phi + B_n \cos^n \phi } \, . \label{frac1}
\ee
By substituting $1-\sin^2\phi$ for all factors of $\cos^2\phi$ in terms of order $n>1$ in the denominator, Eq.~\eqref{frac1} can be written as
\be
\int_0^{2\pi} d\phi \ \frac{\sum_m C_m \sin^m \phi + D_m \cos^m \phi }{\sum_n E_n \sin^n \phi + F_n \cos \phi \sin^n \phi } \, . \label{frac2}
\ee 
Multiplying both the numerator and denominator of Eq.~\eqref{frac2} by $\sum_n E_n \sin^n \phi - \sum_n F_n \cos \phi \sin^n \phi $, gives even-only orders of $\cos\phi$ in the denominator. By repeating $\cos^2\phi \rightarrow 1-\sin^2\phi$, the denominator becomes a polynomial of the variable $\sin\phi$. By factorizing this polynomial using its roots $a_l$ (which are not in the real interval $[-1,1]$), and then splitting the product of fractions to a sum of fractions, the integral over $\phi$ can be written as
\be
 \sum_{l,m,n} G_{l,m,n} \int_0^{2\pi} d\phi \  \frac{e^{im\phi}}{(\sin\phi-a_l)^n} \, . \label{frac3}
\ee
The integrals $R_{l,m,n} = \int_0^{2\pi} d\phi \  \frac{e^{im\phi}}{(\sin\phi-a_l)^n}$ in Eq.~\eqref{frac3} can be evaluated analytically. 

For example, for zero component of the imaginary part of quark self-energy we have the closed from 
\be 
{\rm Im} \{ \boldsymbol \Sigma^0 (z) \} = - \frac{m_q^2 }{4 k} \ \frac{ \Theta(1-z^2)}{2 \xi \sqrt{-\frac{1}{\xi}}\sin\theta_k \sqrt{1-z^2}} \  \Big[ S(a_{+})-S(a_{-})\Big] \, ,
\ee
with
\be
S(a) = -\frac{2 \pi  \sqrt{\frac{a^2}{a^2-1}}}{a} \, , \qquad  \qquad 
a_{\pm} = \frac{z \cot \theta
   _k}{\sqrt{1-z^2}} \pm \csc \theta _k \sqrt{-\ \frac{1}{\xi 
   \left(1-z^2\right)}} \nonumber \, ,
\ee
where $\Theta(x)$ denotes the Heaviside (step) function.

\section{Basis functions $I_n(z)$}
\label{appndxb}

When expanding the self-energy integrals for general $z$ we used the basis functions  
\ba
I_n(z)  &=&   \int_{-1}^{1} dx \ \frac{x^n}{z - x}  \nonumber \\
&=& \frac{  _{2}F_{1}(1,1+n,2+n,1/z) \ + \  (-1)^{n} \ _{2}F_{1}(1,1+n,2+n, -1/z)}{(1+n)z} \, .
\ea
For these basis functions, there is a simple recursion relation
\be
I_{n+1}(z) = z \, I_n(z) - \frac{1 + (-1)^n}{n+1}  \quad ; \quad \text{for\ } n>0 \, .
\ee
A few first orders of these functions are
\ba
I_0(z) &=& \ \log\left(\frac{1+z}{1-z}\right) ,  \\
I_1(z) &=& \ z \log\left(\frac{1+z}{1-z}\right) -2 \, ,  \\
I_2(z) &=& \  z^2 \log\left(\frac{1+z}{1-z}\right) -2z \, .
\ea
At any order, such expressions can also be obtained using the following identity
\be
_{2}F_{1}(1,1+n,2+n,x) = - \frac{n+1}{x^{n+1}} \left[ \log(1-x) + \sum_{m=1}^n \frac{x^m}{m} \right] .
\ee
For large values of $z$, the approximation below can be very useful
\be
I_n(z) \approx  \frac{1}{(n+1)z} \left[  (-1)^n  \left( 1- \frac{n+1}{(n+2)z} \right)  + \left(  1+\frac{n+1}{(n+2)z} \right) \right] .
\ee

\bibliography{im2re}

\end{document}